\documentclass[twocolumn,conference]{IEEEtran}
\usepackage[latin9]{inputenc}
\usepackage{booktabs}
\usepackage{calc}
\usepackage{textcomp}
\usepackage{amsmath}
\usepackage{graphicx}
\usepackage{multirow}

\usepackage[unicode=true,
 bookmarks=true,bookmarksnumbered=true,bookmarksopen=true,bookmarksopenlevel=1,
 breaklinks=false,pdfborder={0 0 0},pdfborderstyle={},backref=false,colorlinks=false]
 {hyperref}
\hypersetup{pdftitle={Your Title},
 pdfauthor={Your Name},
 pdfpagelayout=OneColumn, pdfnewwindow=true, pdfstartview=XYZ, plainpages=false}

\usepackage{todonotes}
\makeatletter

\mathchardef\mhyphen="2D

\usepackage[caption=false,font=footnotesize]{subfig}

\makeatother

\begin{document}

\title{Data Mining with Big Data in Intrusion Detection Systems: A Systematic Literature Review}

\author{\IEEEauthorblockN{Fadi Salo\IEEEauthorrefmark{1}, MohammadNoor Injadat\IEEEauthorrefmark{1},
Ali Bou Nassif\IEEEauthorrefmark{2}\IEEEauthorrefmark{1}, Aleksander Essex\IEEEauthorrefmark{1}}\IEEEauthorblockA{\IEEEauthorrefmark{1}Department of Electrical and Computer Engineering,
The University of Western Ontario, London, ON, Canada \\
Email: \{fsalo, minjadat, aessex\}@uwo.ca}\IEEEauthorblockA{\IEEEauthorrefmark{2}Department of Electrical and Computer Engineering,
University of Sharjah, Sharjah, UAE\\
Email: anassif@sharjah.ac.ae}}

\maketitle
\begin{abstract}
Cloud computing has become a powerful and indispensable technology for complex, high performance and scalable computation. The exponential expansion in the deployment of cloud technology has produced a massive amount of data from a variety of applications, resources and platforms. In turn, the rapid rate and volume of data creation has begun to pose significant challenges for data management and security. The design and deployment of intrusion detection systems (IDS) in the big data setting has, therefore, become a topic of importance. In this paper, we conduct a systematic literature review (SLR) of data mining techniques (DMT) used in IDS-based solutions through the period 2013-2018. We employed criterion-based, purposive sampling identifying 32 articles, which constitute the primary source of the present survey. After a careful investigation of these articles, we identified 17 separate DMTs deployed in an IDS context. This paper also presents the merits and disadvantages of the various works of current research that implemented DMTs and distributed streaming frameworks (DSF) to detect and/or prevent malicious attacks in a big data environment. 
\end{abstract}


\IEEEpeerreviewmaketitle{}

\section{Introduction}
Enterprises, like government, finance, industry, and health care, need strong, reliable cybersecurity solutions \cite{byres2004myths}. Undoubtedly, the challenge has increased dramatically with the big data era from the current implemented solutions, such as traditional intrusion detection systems (IDS) \cite{OSANAIYE2016147}. The substantial recent growth of technological capabilities paves the way to cyber threats that target individuals or enterprises with various malicious attacks. In this threat environment, traditional security protections like firewalls, anti-virus software, and virtual private networks (VPNs) are not always sufficient, and robust mechanisms for detecting intrusions must also be deployed. For that reason, deploying an IDS along with the other traditional security systems is an essential strategy \cite{SANGKATSANEE20112227}. An IDS is automated software used to analyze the network traffic to detect and/or prevent malicious attacks \cite{Axelsson}.  It can be classified by two common methods namely, misuse-based detection (MD) and anomaly-based detection (AD). In the first approach,  classification of the attack is based specifically on known patterns, called signatures \cite{ESTEVEZTAPIADOR20041569}. AD, on the other side, seeks to detect abnormal patterns or behavior, with the key challenge of distinguishing between normal and abnormal patterns with low error \cite{ESTEVEZTAPIADOR20041569}. There are different studies discussing the challenges facing IDS especially with the rise of big data era, which brings up different types of malicious attacks. Accordingly, special tools are required to accommodate the scalability of big data to detect cyber threats. The notion of big data is an umbrella term that encompasses the recent trends toward collection, storage and analysis of enormous volumes of data.  Popular big data management tools include Hadoop, Spark, and Shark.
  
To the best of our knowledge, no prior systematic literature review addresses research questions (RQs) about implemented DMTs in the context of IDS, which has motived this work. In related work, Ashraf and Habaebi \cite{ASHRAF2015112} discussed the threat approaches in the Internet of Things using an autonomic taxonomy. Ali et al. \cite{ALI2015357} summarized the security challenges and vulnerabilities in the mobile cloud computing. Ahmed et al. \cite{AHMED201619} presented the major categories of anomaly detection including statistical, classification, and clustering. They also discussed the dataset challenges of network IDS. Zuech et al. \cite{zuech2015intrusion} reviewed the problem of big heterogeneous data. Injadat et al. \cite{INJADAT2016654} summarized the data mining techniques implemented in social media. Alguliyev and Imamverdiyev \cite{7035946} analyzed the big data applications related to information security challenges and illustrated the research directions at the same domain.  
In this study, we conduct a systematic literature review (SLR) following the framework of Kitchenham and Charters, \cite{kitchenham2007guidelines} exploring research into DMTs in the big data setting for IDS. The study covers the period between Jan 2013 and Dec 2018, evaluates, and analyzes research conducted in this space. This paper is divided into four sections. Section \ref{Sec:Methodology} describes our review methodology. Section \ref{Sec:Results} presents the results and discussion. Section \ref{Sec:Conclusion} includes the conclusion and recommendations for future work.

\section{Methodology\label{Sec:Methodology}}
\begin{figure}
	\begin{centering}
		\includegraphics[scale=0.75]{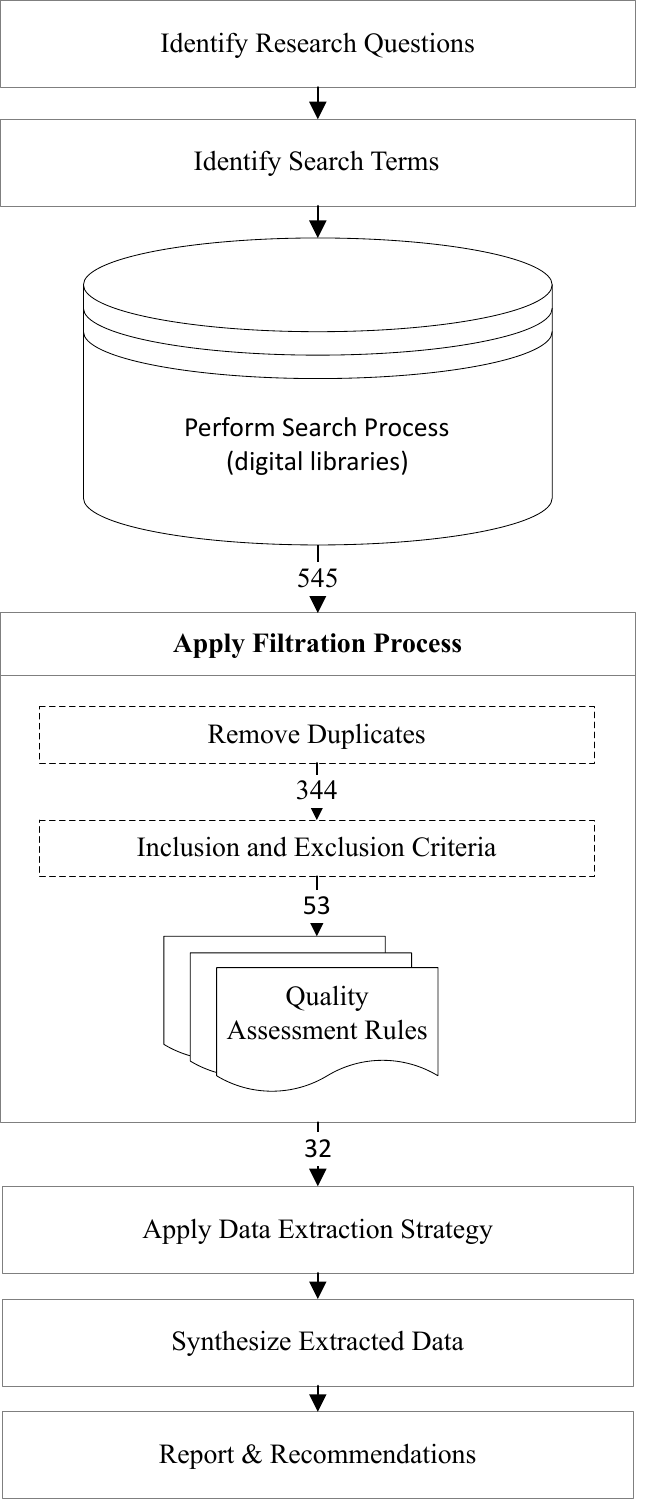}
		\par\end{centering}
	\caption{Proposed framework}
	\label{proposed_framework}
\end{figure}

The systematic methodology of this review consists of three stages \cite{kitchenham2007guidelines}: planning, conducting, and reporting, with  each stage consisting of several sub-stages. Detailed information of this methodology is presented in the following subsections. Fig. \ref{proposed_framework} demonstrates this methodology. 
\subsection{Research Questions (RQs)}
The key element of this study is to analyze and report the empirical research about the DMT with big data in the field of IDS during the desired period. Towards this goal, four RQs were identified: RQ1: Which data mining techniques with big data have been used in IDS? RQ2: What are the strengths and weaknesses of the DMT and DSF employed in the literature? RQ3: What are the evaluation metrics that have been implemented to validate the DMT in IDS? and RQ4: What type of attacks does the IDS in the literature designed for? What are the characteristics of the utilized datasets used to achieve this goal?
\subsection{Study Selection}
This research requires comparative investigation across the relevant articles. The digital libraries used in our review included: IEEE Xplore, ACM, Web of Science, Elsevier, Springer, and Wiley. Our initial search returned 545 articles. Tangential and duplicate articles were then identified during a subsequent filtration step. Next, we applied our inclusion and exclusion criteria to exclude the irrelevant articles. Next, we applied our Quality Assessment Rules (QARs) as a benchmark to evaluate the quality of the candidate articles.
The inclusion criteria consist of the following rules: 1) Include articles that use data mining techniques with big data in IDS. 2) Include comparative studies of data mining techniques with big data in IDS. 3) Include the most recent version of articles (if any). 4) Consider the articles during the publication period between Jan 2013 and Dec 2018. On the other hand, exclusion criteria consisted of these rules: 1) Exclude articles that involve data mining with big data but are not related to IDS. 2) Exclude articles that involve IDS but are not related to data mining with big data. 3) Exclude articles that involve data mining in IDS but are not related to big data. 4) Exclude articles not categorized as peer-reviewed journal or conference papers. After applying the inclusion and exclusion criteria, we obtained 53 journal and conference articles as the primary source of the quality assessment stage.
\subsection{Quality Assessment Rules (QAR)}
As a step to guarantee the quality of the articles, 10 quality assessment questions were identified as follows: QAR1: Is the research purpose/problem clearly defined? QAR2: Are the DMT utilized with big data clearly defined? QAR3: Is the IDS principle clearly described? QAR4: Is the data preprocessing in the experiment well suited for the intended research? QAR5: Is the employed dataset clearly defined? QAR6: Are the big data mining tools clearly pinpointed? QAR7: Are the merits and shortcomings of the employed big data analysis tool, software, or algorithm addressed? QAR8: Is the background of attacks clearly depicted? QAR9: Does the research apply accurate evaluation metrics? and QAR10: Are the experiment results and findings are clearly reported? 

Based on these QARs, we evaluated the 53 articles to assure the richness and suitability of the final outcome.  Each question was graded from 0 to 1 mark, where each QAR was scored as follows: ``Excellent'' = 1, ``Very good'' = 0.75, ``Good'' = 0.5, ``Poor'' = 0.25, ``not answered'' = 0. Accordingly, article with score 5 or more was considered for the data extraction stage. Overall, a total of 32 articles were considered in the data extraction stage. Table \ref{tab:ARTICLES PER SCORE} list the selected articles group by the final score.

\begin{table}
	\centering
	\caption{Articles per score\label{tab:ARTICLES PER SCORE}}
	\begin{tabular}{lcc}
		\hline
		Paper ID& Score & No. of papers 	\\
		\hline
		P1, P6, P8, P26			&	5	&	4 \\
		P2, P9, P12, P15, P27, P29	&	5.5	&	6 \\
		P4, P11, P16, P17, P22	&	5.75&	5\\
		P5, P14, P20, P28, P31	&	6	&	5\\
		P7, P21, P30					&	6.25&	3\\
		P10, P13, P19, P24		&	6.5	&	4\\
		P23						&	6.75&	1\\
		P3, P18, P25			&	7.25&	3\\
		P32						&	8	&	1\\
	\hline
	\end{tabular}
\end{table}
\subsection{Extract and Synthesize Data}
The aim of this stage is to extract information that will answer the RQs. For the sake of quality assurance, each author plays the role of extractor and checker of others? work. To ensure the consistency of the analytical results, common names were used for the same DMT, software, or big-data tool. For instance, Bayesian Network replaced the other related names such as Naive Bayes, Naïve Bayes, or Naïve Bayesian. Different methods were adopted from \cite{kitchenham2007guidelines} to provide evidence of the collected data that are sought to provide answers to the present RQs. We used narrative synthesis (RQ1, RQ3, and RQ4), binary outcomes (All RQs), and reciprocal translation (RQ2) methods since those methods outfit the answers of our RQs. Table \ref{tab:SELECTED ARTICLES } shows the RQs that were answered by each selected article.
\begin{table}
	\centering
	\caption{Selected articles \label{tab:SELECTED ARTICLES }}
	
	\begin{tabular}{lcccccc}
		\hline
		ID&	RQ1&	RQ2&	RQ3&	RQ4&	Year&	Ref 	\\
		\hline
		P1&		1&	1&	1&	0&	2013&	\cite{SuthaharanRef}
\\
		P2&		0&	1&	0&	1&	2014&	\cite{7004484}
\\
		P3&		1&	1&	1&	1&	2014&	\cite{SINGH2014488}
\\
		P4&		1&	1&	0&	0&	2014&	\cite{7034762}
\\
		P5&		1&	1&	0&	1&	2015&	\cite{OginoRef}
\\
		P6&		1&	0&	1&	1&	2015&	\cite{SAIT2015}
\\
		P7&		1&	0&	1&	1&	2015&	\cite{CHEN201610}
\\
		P8&		1&	0&	1&	1&	2015&	\cite{JABEZ2015338}
\\
		P9&		1&	1&	1&	0&	2015&	\cite{7396825}
\\
		P10&	1&	0&	1&	1&	2015&	\cite{7130875}
\\
		P11&	1&	0&	1&	0&	2015&	\cite{7149025}
\\
		P12&	1&	1&	1&	0&	2015&	\cite{7363892}
\\
		P13&	1&	0&	1&	0&	2015&	\cite{7165133}
\\
		P14&	1&	1&	1&	0&	2016&	\cite{solaimani2016online}
\\
		P15&	1&	0&	1&	1&	2016&	\cite{jiang2016novel}
\\
		P16&	1&	1&	1&	0&	2016&	\cite{Rathore2016}
\\
		P17&	1&	1&	1&	0&	2016&	\cite{GUPTA2016824}
\\
		P18&	1&	1&	1&	1&	2016&	\cite{Natesan2017}
\\
		P19&	1&	1&	1&	0&	2016&	\cite{AjabiRef}
\\
		P20&	1&	1&	1&	1&	2016&	\cite{7816982}
\\
		P21&	1&	0&	1&	1&	2016&	\cite{7508144}
\\
		P22&	1&	0&	1&	1&	2016&	\cite{7841864}
\\
		P23&	1&	1&	1&	1&	2017&	\cite{Boukhris2017}
\\
		P24&	1&	0&	1&	1&	2017&	\cite{rachburee2017big}
\\
		P25&	1&	1&	1&	1&	2017& \cite{8125840}
\\
		P26&	1&	1&	1&	1&	2017& \cite{SiddiqueRef}
\\
		P27	&	1&	1&	1&	1&	2018& \cite{GulmezRef}
\\
		P28	&	1&	1&	1&	1&	2018& \cite{DahiyaRef}
\\
		P29	&	1&	1&	1&	1&	2018& \cite{SalmanRef}
\\
		P30	&	1&	1&	1&	1&	2018& \cite{hatef2018hidcc}
\\
		P31	&	1&	1&	1&	1&	2018& \cite{Othman2018}
\\
		P32	&	1&	1&	1&	1&	2018& \cite{KurtRef}
\\		
		\hline
	\end{tabular}
\end{table}

\section{Results and Discussion\label{Sec:Results}}
\subsection{RQ1: Which data mining techniques with big data have been used in IDS?}
\begin{figure}
	\begin{centering}
		\includegraphics[scale=0.48,trim=2.7cm 8.5cm 2.7cm 7cm]{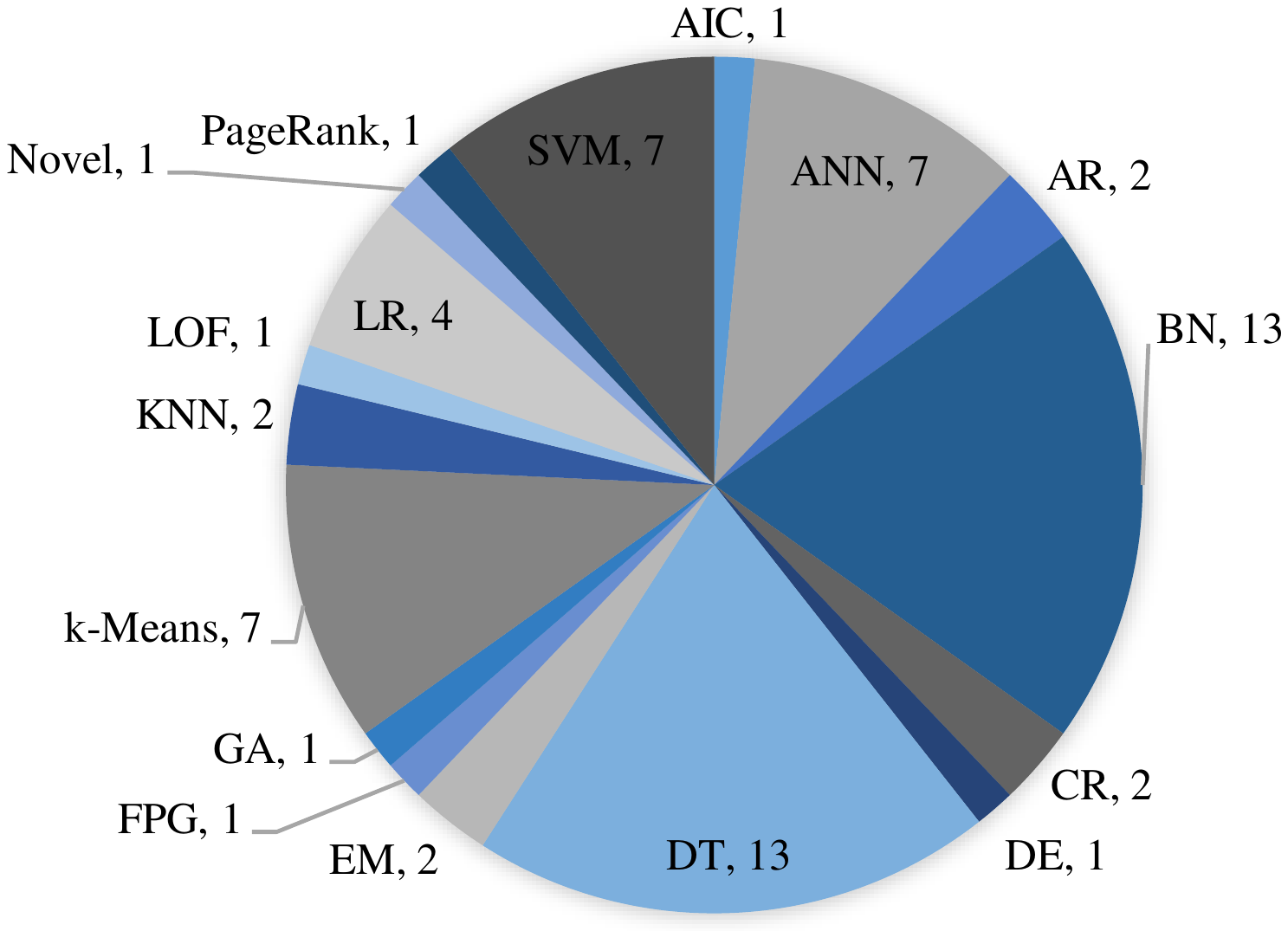}
		\par\end{centering}
	\caption{Utilized data mining techniques}
	\label{Utilized DMT}
\end{figure}

Throughout the selected articles, we have identified 16 DMTs in addition to one novel technique named Unit Ring Machine (URM). The techniques are: Adaptive Incremental Clustering (AIC), Artificial Neural Network (ANN), Association Rule (AR), Bayesian Networks (BN), Conjunctive Rule (CR), Density Estimation (DE), Decision Tree (DT), Expectation-Maximization (EM), FP-Growth (FPG), Genetic Algorithms (GA), \textit{k}-Means, \textit{k}-Nearest Neighbors (\textit{k}NN), Local Outlier Factor (LOF), Logistic Regression (LR), PageRank, Support Vector Machines (SVM), and URM. As shown in Fig. \ref{Utilized DMT}, DT and BN are by far the most widely used techniques in the IDS that suit the nature of big data with a total of 13 papers each. Some experiments propose enhancement features in the traditional DMT to acquire the desirable scalability in machine learning module. A novel technique (URM) was proposed in \cite{SuthaharanRef} which uses the geometric properties of both normal and intrusion network traffic. The experimental results revealed that URM shows high suitability with big data in IDS. 

\begin{figure}
	\begin{centering}
		\includegraphics[scale=0.4,trim=2.7cm 7cm 2.7cm 8cm]{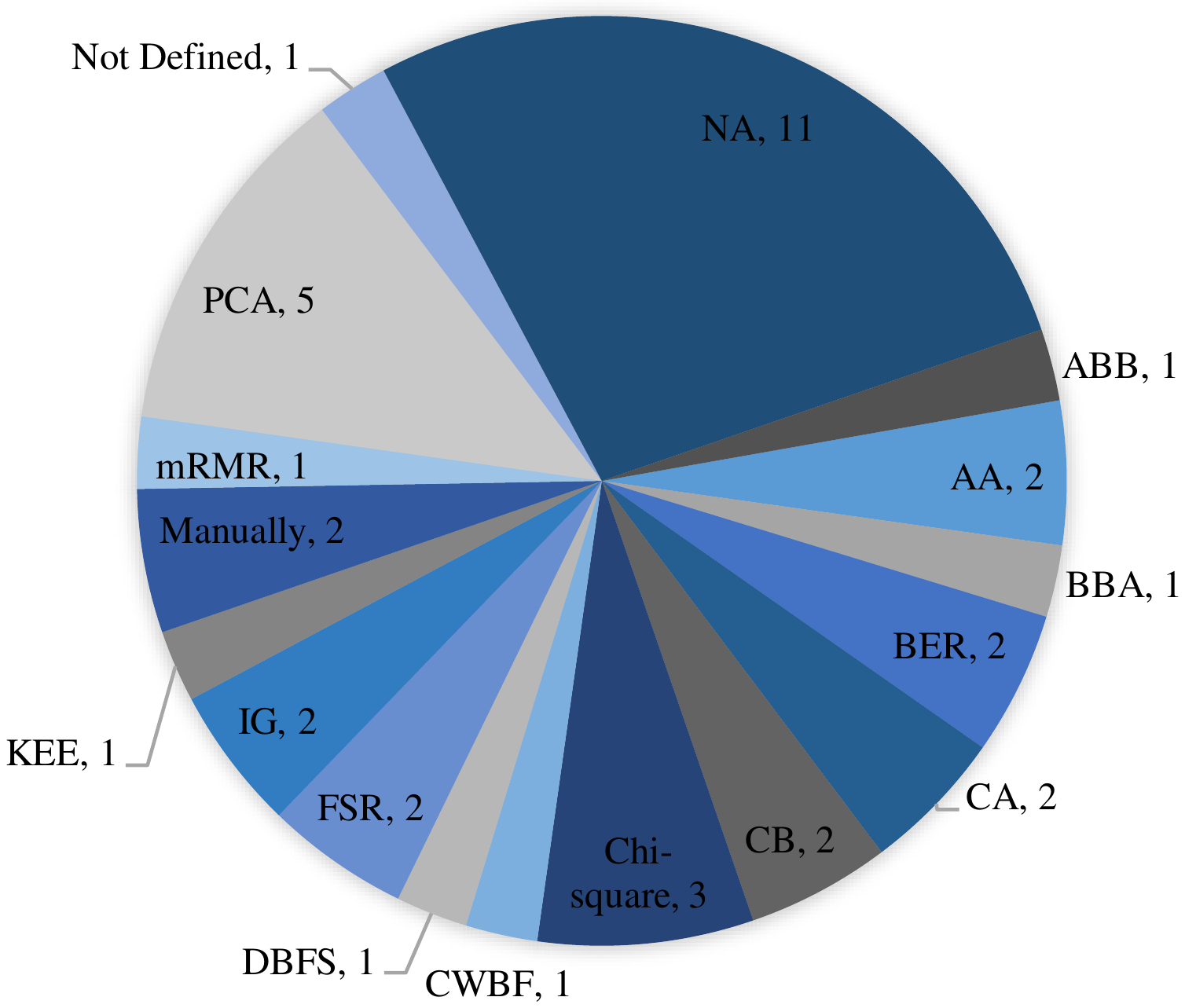}
		\par\end{centering}
	\caption{Feature selection techniques}
	\label{Feature selection techinques}
\end{figure}

Next, we enumerated the DMTs used in the feature selection/extraction (FS) phase since it is a vital step in making classification computationally feasible and reasonable accurate. Fig. \ref{Feature selection techinques} demonstrates 15 DMTs used for FS, which include:  Averaging Approach (AA), Automated Branch-and-Bound (ABB), Binary Bat Algorithm (BBA), Backward Elimination Ranking (BER), Conjunctive Approach (CA), Correlation Based (CB), Chi-square, Connection-Window Based Features (CWBF), Distribution-Based Feature Selection (DBFS), Forward Selection Ranking (FSR), Information Gain (IG), Knowledge Engineering Expert (KEE), Minimum Redundancy Maximum Relevance (mRMR), and Principal Component Analysis (PCA). As shown in Fig. \ref{Feature selection techinques}, PCA is the most popular technique used to reduce the dimensionality. Table \ref{tab:pre-processing techniques} shows the pre-processing techniques applied by  the included articles.

It is worth mentioning, as depicted in Table \ref{tab:pre-processing techniques}, that a big number of experiments did not consider the FS step in their approach, although it is important step to enhance the speed and effectiveness of the proposed models, especially in the big data setting. Section \ref{subsec:RQ2} includes further discussion about the utilized DMTs. 

	\begin{table}
	\centering
	\caption{Utilized pre-processing techniques\label{tab:pre-processing techniques}}
	\scalebox{0.9}{
		\begin{tabular}{p{4.5cm}p{4cm}}
			\hline
			Paper ID& Pre-processing Technique \\
			\hline
			P3, P6, P8, P9, P15, P16, P17, P18, P19, P22, P23, P25, P30	&FS\\
			P1, P12, P29	&Normalization, FS\\
			P13, P26, P31	&Encoding, Normalization, FS\\
			P11, P24	&Encoding, Data cleansing, FS\\
			P14	&Normalization, Data cleansing\\
			P7	&Normalization\\
			P32	&Encoding\\
			P2, P4, P5, P10, P20, P21, P27, P28	&NA\\
			\hline
	\end{tabular}}
\end{table}
\subsection{RQ2: What are the strengths and weaknesses of the DMT and DSF employed in the literature?\label{subsec:RQ2}}	 
This RQ provides a rich source of information about the utilized DMTs and big data tools in the literature. We counted 9 different DSFs including: Spark, Hadoop, Kafka, Storm, Azure, Eucalyptus, HAMR, Jubatus, and Mahout. Among the included articles, it has been observed that Apache Spark outperforms Hadoop and other big data distributed streaming frameworks in term of speed, data streaming, and modifying the data in real time. Tables \ref{tab:DMT STRENGTHS AND WEAKNESSES} \& \ref{tab:DSF STRENGTHS AND WEAKNESSES} present the experiments and knowledge outcomes, which form a potential source and guideline for future research in this topic area. 
\subsection{RQ3: What are the evaluation metrics that have been implemented to validate the DMT in IDS?}	
\begin{figure}
	\begin{centering}
		\includegraphics[scale=0.45,trim=2.9cm 7cm 2.7cm 8cm]{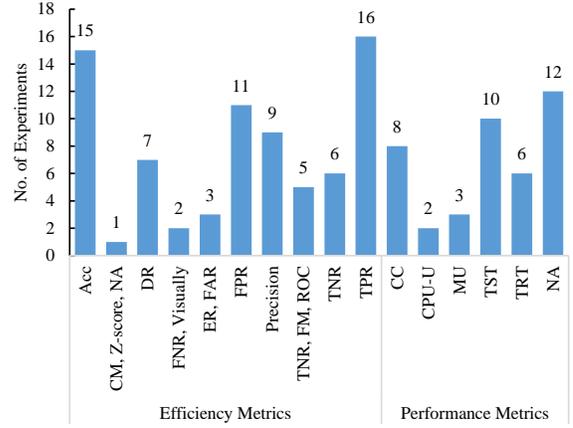}
		\par\end{centering}
	\caption{Evaluation metrics}
	\label{Evaluation metrics}
\end{figure}

The aim of this RQ is to enumerate the most popular evaluation metrics utilized by the respective IDSs. Overall, 19 approaches were identified in the related literature, detailed explanation about the techniques are discussed in \cite{munaiah2016intrusion}. We categorized our findings into two metrics: efficiency and performance. The efficiency metric includes 16 methods: Accuracy (Acc), Confusion Matrix (CM), Detecting Rate (DR), Error Rate (ER), False Alarm Rate (FAR), F-Measure (FM), False Negative Rate (FNR), False Positive Rate (FPR), Precision, ROC Curve (ROC), True Negative Rate (TNR), True Positive Rate (TPR), Z-score, and Visual method.  Performance metrics, on the other side, are: Computational Complexity (CC), CPU Utilization (CPU-U), Memory Utilization (MU), Testing Time (TST), Training Time (TRT). Fig. \ref{Evaluation metrics} shows that TPR, Acc, and FPR are the most commonly used for efficiency measurement, while the popular performance measurements are TST, CC, and TRT.
\begin{figure}
	\begin{centering}
		\includegraphics[scale=0.45,trim=2.7cm 7cm 2.7cm 8cm]{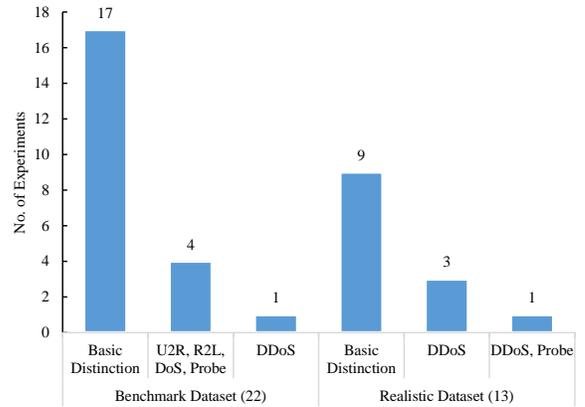}
		\par\end{centering}
	\caption{Attacks vs datasets}
	\label{Attacks vs datasets}
\end{figure}
Even though evaluation metrics are considered vital to measure and evaluate the performance of IDSs, some papers did not validated their results sufficiently insofar as they relied on a single evaluation metric as a performance indicator, which is known to generate misleading results in some cases \cite{chawla2002smote}. For instance, P2, P9, P11, and P13 reported either Acc, ER, or FPR in their studies. In the same context, we observed 12 articles did not consider the performance metrics utilization at all.
\subsection{RQ4: What type of attacks does the IDS in the literature designed for? What are the characteristics of the utilized datasets used to achieve this goal? }
Attacks can be classified into four major types: Denial of Service (DoS), Remote to User (R2L), User to Root (U2R), and Probing. Based on the data extracted, we classified the results either based on realistic datasets captured from blogs, profiles, and network traffic or benchmark datasets including DARPA, KDD Cup 99, NSL-KDD, ISCX-UNB, UNSW-NB15, and CAIDA. Detailed information about these attacks and datasets are discussed in \cite{7456894}. We noticed, as shown in Fig. \ref{Attacks vs datasets}, the majority of the experiments are designed to provide binary classification of behaviors, for example, as being either normal or malicious. Furthermore, none of the real-time IDS were able to classify intrusions into the four major types. It is worth to mention that a significant number of the experiments are based the DARPA/KDD datasets which they considered obsolete and does not capture contemporary attack types.
\begin{table*} 
	\caption{Data mining strengths \& weaknesses\label{tab:DMT STRENGTHS AND WEAKNESSES}}
	\centering
	\scalebox{0.9}{
		\begin{tabular}{p{1cm}p{1cm}p{12cm} p{2cm}}
			\hline
			\multicolumn{2}{l}{Technique}                   & Strength                                                                                                                                                                                                                                      & Ref   \\
			\hline                
			AR                      & Cons                  & Apriori: High memory \& time usage.                                                                                                                                                                                                           & P15                   \\
			URM                     & Pros                  & Efficient with big data. Minimize the noise boundary between network traffic. Adjustable widths for representation learning.                                                                                                                  & P1                    \\
			LOF                     & Pros                  & Superior in detecting outliers.                                                                                                                                                                                            & P5                    \\
			\multirow [t]{3}{*}{SVM}    & Pros                  & Work efficiently with multi-dimensional complex data.                                                                                                                                                                                       & P3, P17, P25               \\
			& \multirow[t]{2}{*}{Cons}                  & Less efficiency in decision making \& model  building.                                                                                                                                                                                        & P16                   \\
			&                       & Computationally expensive, especially with big datasets.
			& P31                    \\		
			\multirow [t]{4}{*}{BN}     & \multirow[t]{3}{*}{Pros} & Efficient with large and complex data.                                                                                                                                                                                                                  & P16, P17, P25              \\
			&                       & Handle missing or noisy data efficiently.                                                                                                                                                                                                     & P18                   \\
			&                       & Well-perform with big data classification and irrelative attributes.                                                                                                                                                                          & P2, P7, P17, P32           \\
			& Cons                  & Not efficient in decision-making.                                                                                                                                                                                                             & P16                   \\
			\multirow [t]{2}{*}{\textit{k}-NN}   & Pros                  & Flexibility with multi-dimensional data. Useful for nonlinear data data.                                                                                                                                    & P3                    \\
			& Cons                  & Computationally expensive, especially with big data. Sensitive to irrelevant features.                                                                                                                                                                                                       & P3                    \\
			\multirow[t]{3}{*}{DT}     & \multirow[t]{2}{*}{Pros} & REPTREE: efficient in decision making and model building.                                                                                                                                                                                     & P16                   \\
			&                       & Suitable  for high-dimensional data and irrelevant descriptors.                                                                                                                                                                               & P3                    \\
			& Cons                  & Low prediction accuracy, especially with unexpected events.                                                                                                                                                                                                                       & P3                    \\
			\multirow [t]{5}{*}{ANN}    & \multirow[t]{5}{*}{Pros} & Flexibility  with multi-dimensional data during model building.                                                                                                                                                                               & P3                    \\
			&                       & GHSOM: Efficient at clustering samples with high-dimensionality. Adaptive while creating hierarchical clusters.                                                                                                                                  & P4                    \\
			&                       & Learning Vector Quantization (LVQ): Computationally not expensive.                                                                                                                                  & P30                    \\
			& \multirow [t]{2}{*}{Cons} & Inefficient with low-frequent attacks (U2R \& R2L). Poor detection  stability.                                                                                                                                                                & P8                    \\
			&                       & Computationally more expensive than other traditional algorithms.                                                                                                                                                                                                         & P3                    \\
			NOF                     & Pros                  & Stable detection performance.                                                                                                                                                                                                                 & P8       \\   
			\textit{k}-Means & Pros& \textit{k}-Medoids: Reduces the cost between non medoids and medoids of the cluster. & P28 \\         
			\hline
	\end{tabular}}
\end{table*}
\begin{table*}
	\caption{Distributed streaming frameworks strengths \& weaknesses\label{tab:DSF STRENGTHS AND WEAKNESSES}}
	\centering
	\scalebox{0.9}{
		\begin{tabular}{p{1cm}p{1cm}p{12cm} p{2cm}}
			\hline
			\multicolumn{2}{l}{Technique}                   & Strength                                                                                                                                                                                                                                      & Ref   \\
			\hline                
			\multirow [t]{4}{*}{Spark}  & \multirow[t]{3}{*}{Pros} & Speed: high data processing. Dynamic: Perform stream computations as a series of micro-batch jobs. Real-time stream processing: Spark streaming can handle and process the present and real-time data. Usability: Simple cluster setup procedure. Can handle bottleneck problem available with MapReduce. & P14, P17, P25              \\
			&                       & Efficiency: With DAG execution engine, Spark is faster than Hadoop in memory and on the  disk. Support multiple languages such as Python, R, and Java.                                                                                                                                                                                                    & P7,P14, P17, P25, P27, P31, P32           \\
			&                       & Flexibility: Can run independently or integrated with different platforms. Suitable for iterative machine learning. Allow changing data in real-time. Fault tolerance: designed to handle the failure at any node, which reduces the data losing to zero.                                                           & P7, P17               \\
			& Cons                  & Less number of algorithms: Spark includes few standard DMTs.                                                                                                                                                                                                                    & P14                   \\
			
			\multirow [t]{3}{*}{Kafka}  & \multirow[t]{3}{*}{Pros} & Compatible with other platforms such as Storm and Spark. High-throughput: Kafka has the capability of handling high-volume \& high-velocity data.                                                                                                                                                                                     & P14                   \\
			&                       & Load balancing: Work with multiple-data sources at the same time. Fault tolerance: Resistant to machine/node failure.                                                                                                                                                    & P11, P14, P27              \\
			&                       & Durability: Data streamed into the Kafka framework can be persistent on disk.                                                                                                                                                                           & P11                   \\
			Storm                   & Cons                  & Less number of algorithms: includes few standard DMTs.                                                                                                                                                                                                                    & P14                   \\
			\multirow [t]{6}{*}{Hadoop} & \multirow[t]{2}{*}{Pros} & Better security than Spark.                                                                                                                                                                                                                   & P7, P23               \\
			&                       & Scalability: highly scalable storage platform. Can be operate over large number of nodes/machines involving terabytes of data. Resilient to failure: able to resume crashed process.                                                                                                                                                                                                & P7, P18, P19, P21, P23, P29 \\
			& \multirow[t]{4}{*}{Cons} & Not Fit for Small Data: While Hadoop has high capacity design  towards big data platforms, it is therefore not suitable to random handling of small files.                                                                                                                                                                                                               & P14, P5               \\
			&                       & Real-time stream processing: Not suitable for real-time analysis.                                                                                                                                                                                                           & P5, P11               \\
			&                       & Stability issues: Exposed to hardware failures.                                                                                                                                                                                                                  & P7                    \\
			&                       & Security concerns: Lacks of monitor users' activities  monitoring                                                                                                                                                                                                                 & P11                   \\
			Mahout                  & Cons                  & Good for batch processing only.                                                                                                                                                                                                               & P14                   \\
			Jubatus                 & Pros                  & Effective in deep analysis and real-time distribution.                                                                                                                                                                                        & P5                    \\
			HAMR                    & Pros                  & Support MapReduce programming. Faster than Mahout \& Hadoop. Less latency over Storm. Consumes less memory resource than other platforms.                                                                       & P2                    \\
			\hline
	\end{tabular}}
\end{table*}		

\begin{table*}
	\centering
	\caption{Comprehensive summary of the included articles\label{tab:COMPREHENSIVE SUMMARY OF THE INCLUDED ARTICLES}}
	\scalebox{0.9}{
		\begin{tabular}{p{0.6cm}p{8.3cm}p{8.3cm}}
			\hline
			Ref& Main Contribution & Future Work 	\\
			\hline
			P1&	Utilize the geometric patterns of the network traffic variables	&Consider multiple-domain, representation-learning 
			with knowledge-transfer \& class-separate objectives\\
			P2&	Propose B-dIDS using HAMR to identify attacks	&Compare B-dIDS against other frameworks \\
			P3&	Capture packets in quasi real-time mode. Propose a distributed framework to characterize flow statistics of the packets. Peer-to-peer detection	& Detect suspicious stealthy communication \\
			P4&	Utilize GHSOM to discover event patterns to discover attacks	&Include more features such as event, protocol and port \\
			P5&	Implement LOF on Jubatus for cyber attack& \\	
			P6&	New method on TCP/IP stack for internal and external threats&	\\
			P7&	Develop a system to secure critical infrastructure systems. Monitor network activities & Enhance the security by adopting proactive \& reactive defense strategies. Handle streaming data \\	
			P8&	Detect anomaly by the neighborhood outlier factor 
			& Employ the proposed approach for different distance computations\\
			P9&	Illustrate the benefit of using structured ML for learning intelligent security in the perfect \& imperfect domain knowledge and detection intrusions	& Conduct further empirical study to identify the benefits of ontology \& rule learning on reducing error rates\\
			P10&	Propose a security enforced architecture for mobile computing	&More security for mobile and non-mobile clouds\\
			P11& Present IDS able to capture, processing, and analysis real-time traffic data	& Improve the ML \& optimization algorithms for better real-time detection. Include visualization tools\\
			P12&	Consider user behavior profiles for anomaly detection&	Inspect new DM technique using applications such as HBase and Pig\\
			P13&	Improve pre-existing SIEM using integrated SAP HANA&	Include more DM technique in R to improve detection techniques\\
			P14&	Propose a distributed novel anomaly detection framework 
			over a large number of virtual machines& 	Consider different DM \& statistical techniques. Utilize other performance metrics. Develop ensemble approach to reduce noise influence. 
			Use different heterogeneous data from\\
			P15& Present a novel mechanism to detect intrusion based on honeypot log similarity analysis \& DM techniques to predict \& stop any suspicious activities prior occurrence&	Provide a multilayer defense \\ 
			P16& Propose ultra high-speed real-time IDS consists of: capturing, filtration, load balancing, Hadoop, and decision-making layers &	 \\
			P17&	Investigate different FS and ML teachings over Apache Spark	& \\
			P18& Propose a scalable IDS that uses parallel BBA technique for FS \& distributed classification model 	& \\
			P19& Perform the classification process in an uncertain environment within the averaging approach& Utilize belief DT 
			using the conjunctive approach \\
			P20& Investigate the performance of the TBS in the big data environment. Combine TBS with random subset selection (RSS)	&\\
			P21& Investigate various ML algorithms in big data environment. Investigate the performance of two versions of Hadoop.
			& 	Explore different ML algorithms \& big data tools \\
			P22& Propose ultra high-speed real-time IDS consists of: capturing, filtration, load balancing, processing, and decision-making layers & \\	
			P23&	Utilize belief DT for big data in an uncertain environment&	Extent the work to classify new uncertain instances\\
			P24&	Utilize mutual information and chi-square for FS on MS azure&	Analyse network traffic in a real-time mode\\
			P25& Propose a big data framework to investigate various ML algorithms. Apply FS \& dimensionality reduction methods&	Design hybrid approach. Provide attack classification \\
			P26&	Handle large volume of network traffic in real-time mode. Propose a novel approach to select optimal feature subsets&	Utilize additional classification techniques\\
			P27&	Detect anomalies in near real-time mode&	Adopt deep reinforcement learning based model. Perform experiment on real-time stream data. Provide attack classification\\
			P28&	Integrate parallel \textit{k}-medoid clustering with \textit{k}-NN&	 \\
			P29&	Propose a framework uses LVQ \& PCA&	 \\
			P30&	Propose a hybrid approach to detect internal \& external attacks. Include both signature-based \& anomaly-based models& 	 \\
			P31&	Conduct a comparison between chi-SVM \& chi-logistic regression& Provide attack classification	 \\
			P32&	Provide a comparison among different DM technique as a guide for big data researchers& Apply features selection. Detect novel attacks. Utilize different datasets	 \\
			\hline
	\end{tabular}}
\end{table*}

\subsection{Further discussion}	
The discussion in this section pinpoints the contributions and future works of the relevant studies in this SLR. The comprehensive summary in Table \ref{tab:COMPREHENSIVE SUMMARY OF THE INCLUDED ARTICLES} draws information from multiple experiments to form a pool of rich source for future researches.


\section{Conclusions\label{Sec:Conclusion}}
This research explored the relevant empirical studies published in journals and conferences to investigate the utilized DMTs with big data in IDS domain. We manually examined an initial 545 papers, which we eventually reduced to include 32 relevant papers. The findings of this work provide an integrated and unified view of the extracted data to answer the proposed RQs. The conclusion of this SLR is summarized as follows:
\begin{itemize}
\item	RQ1: The most commonly used DMT with big data in IDs are: DT, BN, \textit{k}-Means, ANN, and SVM. Likewise, PCA was the most used technique in FS. Furthermore, a large fraction of papers did not consider FS in their approach, which considered an essential step to improve the overall performance.
\item	RQ2: Hadoop was the most popular DSF with a frequency of 12 experiments. However, other big data tools, such as Spark reported better performance. We also provide an additive summary that highlights strengths and weaknesses of the used big data tools, software, or mining algorithms in the field of IDS.
\item	RQ3: We enumerated the evaluation metrics in terms of efficiency and performance. The most used efficiency metrics are TPR, Acc, and FPR. On the other hand, TST, CC, and TRT were reported more frequently as a performance metrics. In addition, we observed inconsistencies that exist in the evaluation metrics used in IDS.
\item	RQ4: We noticed that most of the experiments have classified attacks as either normal or malicious. In the same scenario, no experiments were able to fully identify the major attack types in real-time mode. We also found 66\% of the experiment were conducted based on outdated dataset (DARPA), which may degrade IDS efficiency.
\end{itemize}
The area of big data analytics in IDS is still in its initial stages, and we anticipate major advances on the horizon. Thus, we recommend that more adaptive DMT be integrated with available big data tools. Moreover, additional effort is required to design real-time IDSs to fully classify several types of attack. Another significant point is that Spark outperforms other popular big data tools such as Hadoop, Storm, and Jubatus. In the light of such a scenario, it is recommended to consider Spark for future research with regards to IDS.
 
\small
\bibliographystyle{IEEEtran}
\bibliography{bibfile}

\end{document}